\documentclass[conference]{IEEEtran}
\usepackage{fancyhdr}

\IEEEoverridecommandlockouts
\usepackage[utf8]{inputenc}
\usepackage{amsthm,amsmath,amssymb,amsfonts,mathrsfs}
\usepackage{color}
\usepackage{cancel}
\usepackage[ruled]{algorithm}
\usepackage[noend]{algpseudocode}
\algrenewcommand\algorithmicindent{1.0em}
\algrenewcommand\algorithmicrequire{\textbf{Input:}}
\algrenewcommand\algorithmicensure{\textbf{Output:}}
\usepackage{setspace}
\let\Algorithm\algorithm
\renewcommand\algorithm[1][]{\Algorithm[#1]\setstretch{1.2}}
\algdef{SN}{When}{EndWhen}[1]{\textbf{when} \(\mbox{#1}\) \textbf{do}}%
\theoremstyle{plain}

\DeclareMathOperator*{\ve}{{\mathbf{v}}}
\DeclareMathOperator*{\veq}{{\hat{\mathbf{v}}}}

\usepackage{mathtools}
\DeclarePairedDelimiter\floor{\lfloor}{\rfloor}
\newcommand{\eq}[1]{Eq.~\eqref{#1}}
\DeclareMathOperator*{\x}{{\boldsymbol{\theta}}}

\title{Q-SHED: Distributed Optimization at the Edge\\ via Hessian Eigenvectors Quantization}

\author{\IEEEauthorblockN{Nicol\`o Dal Fabbro$^\ddag$, Michele Rossi$^\ddag$, Luca Schenato$^\ddag$, Subhrakanti Dey$^{\dag}$}
	\IEEEauthorblockA{\textit{$^\ddag$Department of Information Engineering, University of Padova}, 
		Padova, Italy \\
		\IEEEauthorblockA{\textit{$^{\dag}$Department of Electrical Engineering, Uppsala University}, Uppsala, Sweden}
		}}
\date{\today}
\DeclareMathOperator{\tr}{tr}
\DeclareMathOperator{\diag}{diag}

\begin{document}
\thispagestyle{fancy}
\maketitle
\begin{abstract}
Edge networks call for communication efficient (low overhead) and robust distributed optimization (DO) algorithms. These are, in fact, desirable qualities for DO frameworks, such as federated edge learning techniques, in the presence of data and system heterogeneity, and in scenarios where inter-node communication is the main bottleneck. Although computationally demanding, Newton-type (NT) methods have been recently advocated as enablers of robust convergence rates in challenging DO problems where edge devices have sufficient computational power. Along these lines, in this work we propose \mbox{Q-SHED}, an original NT algorithm for DO featuring a novel bit-allocation scheme based on incremental Hessian eigenvectors quantization. The proposed technique is integrated with the recent SHED algorithm, from which it inherits appealing features like the small number of required Hessian computations, while being bandwidth-versatile at a bit-resolution level. Our empirical evaluation against competing approaches shows that Q-SHED can reduce by up to 60\% the number of communication rounds required for convergence.
\end{abstract}
\section{Introduction}\thispagestyle{fancy}
Solving distributed optimization problems in a communication-efficient fashion is one of the main challenges of next generation edge networks~\cite{EdgeAI}. In particular, much attention is being turned to distributed machine learning (ML) settings and applications, and to the distributed training of ML models via \emph{federated learning} (FL)~\cite{GENERAL_FL}. FL is a distributed optimization (DO) framework motivated by the increasing concerns for data privacy at the user end, and by the convenience of performing distributed processing in \mbox{multi-access} edge computing (MEC) networks. However, DO is particularly challenging in federated edge learning (FEL) scenarios where communication occurs over unpredictable and heterogeneous wireless links~\cite{FEL}. To tackle these challenges, major research efforts have been conducted in recent years~\cite{WalidSaad, UVeQFed, ClaBat}. A common assumption in FEL is that edge devices are equipped with sufficient computing capabilities. Hence, Newton-type (NT) methods, although computationally demanding, have been recently advocated to improve the convergence rate of distributed optimization, while significantly reducing its communication overhead~\cite{DONE, BennisAnalog}. Communication efficient distributed NT (DNT) algorithms like GIANT~\cite{GIANT}, and DONE~\cite{DONE} have shown promising results in configurations with i.i.d. data distributions among devices, but underperform when applied to \mbox{ill-conditioned} problems and heterogeneous data configurations~\cite{NDF_Newton}, which are scenarios of major practical relevance. Some works, like FedNL~\cite{FedNL} and SHED (sharing Hessian eigenvectors for distributed learning)~\cite{NDF_Newton} have been recently proposed to robustify FL in the presence of non i.i.d. data distributions, system heterogeneity and \mbox{ill-conditioning}. A DNT method with \mbox{over-the-air} aggregation has been studied in~\cite{BennisAnalog}. Quantized Newton (QN)~\cite{QN} has investigated the convergence properties of the distributed Newton method when the Hessian matrix is quantized. However, QN entails a communication load proportional to $O(n^2)$, where $n$ is the problem dimensionality, while a linear \mbox{per-iteration} communication complexity of $O(n)$ is desirable. 

In this paper, we present Q-SHED, a new algorithm that extends the recently proposed SHED~\cite{NDF_Newton} via a novel bit-allocation scheme based on incremental Hessian eigenvector quantization. In particular, our main contributions are:
\begin{itemize}
	\item We propose an original \mbox{bit-allocation} scheme for Hessian approximation based on uniform scalar dithered quantization of Hessian eigenvectors, to improve the efficiency of second-order information transmission in a DNT method.
	\item We integrate our \mbox{bit-allocation} scheme with the recently proposed SHED technique~\cite{NDF_Newton}, obtaining a new approach, Q-SHED, based on incremental dithered quantization of Hessian eigenvectors. Q-SHED has a communication complexity of $O(n)$ (inherited by SHED) and handles per-iteration heterogeneity of communication channels of the different edge computers involved in the optimization problem at a bit-resolution (per vector coordinate) level.
	\item We evaluate Q-SHED on two datasets assessing its performance in a standard distributed optimization setup, as well as in a scenario where the transmission quality of communication links randomly fluctuates over time according to a Rayleigh fading model (a popular model for wireless channels). With respect to competing solutions, Q-SHED shows convergence speed improvements of at least 30\% in a non-fading scenario and of up to 60\% in the Rayleigh fading case.
\end{itemize}

\section{Distributed optimization framework}\label{sec:sysModel}
We consider the typical DO framework where $M$ machines communicate with an aggregator to cooperatively solve an empirical risk minimization problem of the form \vspace{-0.15cm}
\begin{equation}\label{eq:minimizationPbm}\vspace{-0.15cm}
	    \min_{\boldsymbol{\theta}} f(\boldsymbol{\theta}) := \frac{1}{N}\sum_{d = 1}^{M}N_df^{(d)}(\boldsymbol{\theta}),
\end{equation}
where $\boldsymbol{\theta} \in \mathbb{R}^{n}$ is the optimization variable, $N_d$ is the number of data samples of the $d$-th machine and $N = \sum_{d = 1}^{M}N_d$. For the convergence analysis of the algorithm, we make the following standard assumption on the cost function $f$:\vspace{-0.1cm}
\newtheorem{assumption}{Assumption}
\begin{assumption}\label{ass:localLCont}
	Let $\mathbf{H}(\boldsymbol{\theta}) := \nabla^2f(\boldsymbol{\theta})$ be the Hessian matrix of the cost $f(\x)$. $f(\x)$ is twice continuously differentiable, smooth, strongly convex and $\mathbf{H}(\boldsymbol{\theta})$ is Lipschitz continuous.
\end{assumption}\vspace{-0.3cm} 
\subsection{Distributed Newton method}
The Newton method to solve (\ref{eq:minimizationPbm}) is:\vspace{-0.1cm}
\begin{equation*}
	\boldsymbol{\theta}^{t+1} = \boldsymbol{\theta}^{t} - \eta_t\mathbf{H}_t^{-1}\mathbf{g}_t,\vspace{-0.1cm}
\end{equation*}
where $t$ denotes the $t$-th iteration, $\mathbf{g}_t = \mathbf{g}(\boldsymbol{\theta}_t) = \nabla f (\boldsymbol{\theta}^t)$,  $\eta_t$ and $\mathbf{H}_t = \nabla^2 f (\boldsymbol{\theta}^t)$ denote the gradient, the step size and the Hessian matrix at iteration $t$, respectively. In the considered DO scenario, we have that:\vspace{-0.15cm}
\begin{equation}\vspace{-0.15cm}
	\mathbf{H}_t = \frac{1}{N}\sum_{d = 1}^{M}N_{d}\mathbf{H}_t^{(d)}, \ \ \mathbf{g}_t = \frac{1}{N}\sum_{d = 1}^{M}N_{d}\mathbf{g}_t^{(d)},
\end{equation}
where $\mathbf{H}_t^{(d)} = \nabla^2f^{(d)}(\boldsymbol{\theta}^t)$ and $\mathbf{g}_t^{(d)} = \nabla f^{(d)}(\boldsymbol{\theta}^t)$ denote local Hessian and gradient of the local cost $f^{(d)}(\boldsymbol{\theta}^t)$ of machine $d$, respectively. 
To get a Newton update at the aggregator, in a FL setting one would need each agent to transfer the matrix $\mathbf{H}_t^{(i)}$ of size  $O(n^2)$ to the aggregator at each iteration, whose communication cost is considered prohibitive in many practical scenarios, especially when $n$ is large. To deal with communication constraints, while still exploiting second-order information, DNT methods use Hessian approximations:\vspace{-0.15cm}
\begin{equation}\label{eq:qN-update}\vspace{-0.15cm}
	\boldsymbol{\theta}^{t+1} = \boldsymbol{\theta}^t - \eta_t\hat{\mathbf{H}}_t^{-1}\mathbf{g}_t,
\end{equation}
where $\hat{\mathbf{H}}_t$ is an approximation of $\mathbf{H}_t$.\vspace{-0.15cm}
\subsection{The SHED algorithm}\vspace{-0.15cm}
In this paper, we propose a DNT approach built upon SHED~\cite{NDF_Newton}, a DNT algorithm for FL designed to require few Hessian computations by FL workers, that efficiently shares (low communication overhead) second-order information with the aggregator, see~\cite{NDF_Newton} for a detailed description. SHED exploits a full-rank approximation of the workers' Hessians by sending to the aggregator the most relevant eigenvalue-eigenvector pairs (EEPs) of the local Hessian, along with a local approximation parameter. Approximations are incrementally improved across iterations, as machines send additional EEPs to the aggregator. By doing so, the Hessian is computed only sporadically and outdated versions of it are used to incrementally improve the convergence rate. Under Lipschitz Hessians, strong convexity and smoothness assumptions, SHED has super-linear convergence.\vspace{-0.15cm}
\subsection{Q-SHED: Hessian eigenvectors quantization}\vspace{-0.08cm}
Let $\mathbf{H} = \mathbf{V}\boldsymbol{\Lambda} \mathbf{V}^\top$, be the eigendecomposition of a machine (edge computer) Hessian matrix, with $\boldsymbol{\Lambda} = \diag(\lambda_{1}, ..., \lambda_{n})$, where $\lambda_{k}$ is the eigenvalue corresponding to the $k$-th unitary eigenvector,  $\mathbf{v}_{k}$. In general, the Hessian is a function of the parameter $\boldsymbol{\theta}$, but here we omit this dependence for ease of notation. We always consider eigenvalues ordered so that \mbox{$\lambda_{1} \geq \lambda_{2}\geq  ... \geq \lambda_{n}$}.
In SHED, a machine shares with the aggregator a parameter $\rho_q$ together with $q$ EEPs, allowing for a full-rank $(q, \rho_q)$-approximation of its Hessian, of the form
\begin{equation}\vspace{-0.15cm}
	\mathbf{H}_{q, \rho_q} = \sum_{i = 1}^{q}(\lambda_i - \rho_q)\mathbf{v}_i\mathbf{v}_i^{\top} + \rho_q \mathbf{I} = \mathbf{V}\boldsymbol{\Lambda}_{\rho_q}\mathbf{V}^{\top},
\end{equation}
where $\mathbf{V} := [\ve_1, ..., \ve_n]$, $\boldsymbol{\Lambda}_{\rho_q} := \diag(\lambda_1, ..., \lambda_q, \rho_q, ..., \rho_q) \in \mathbb{R}^{n\times n}$. In the original SHED algorithm, eigenvectors are transmitted exactly (up to machine-precision). Differently, we here design a quantization scheme for the eigenvectors, obtaining a quantized approximation of the Hessian of the form:\vspace{-0.1cm}
\begin{equation}\label{eq:HessQDef}\vspace{-0.1cm}
	\hat{\mathbf{H}}_{q, \rho_q}(b_1, ..., b_{q}) = \sum_{i =1}^{q}(\lambda_i - \rho_q)\hat{\mathbf{v}}_i(b_i)\hat{\mathbf{v}}_i(b_i)^{\top} + \rho_q \mathbf{I},
\end{equation}
where we denote by $\hat{\mathbf{v}}_i(b_i)$ the $i$-th eigenvector, quantized with $b_i$ bits per vector element. As in~\cite{NDF_Newton}, we fix $\rho_q = \lambda_{q+1}$. We design the quantization scheme so that if an eigenvector $\mathbf{v}_i$ is quantized and transmitted, then at least one bit is assigned to each of its components. The vectors to which no bit is assigned are all set equal to zero, i.e., $\veq_i(0) = \boldsymbol{0}$. We assume that, as in typical machine learning problems, $n \gg 1$. Hence, we design the quantization scheme such that the approximation parameter $\rho_q$ and the eigenvalues $\{\lambda_i\}$ are not quantized and are transmitted exactly (up to machine precision).

\section{Optimal quantization of eigenvectors}\label{sec:optiomalQ}
We formulate the design of the quantization scheme as a bit allocation problem, exploiting the specific structure of the Hessian. In particular, as, e.g., in~\cite{UVeQFed}, we consider dithered quantization, so that we can model the quantization error as a uniformly distributed (in the lattice) zero mean additive random noise. Let $\mathbf{v}_i$ be a Hessian eigenvector and let $\hat{\mathbf{v}}_i = \hat{\mathbf{v}}_i(b_i)$ be the same eigenvector quantized with $b_i$ bits per vector coordinate (to improve readability, the dependence on $b_i$ is omitted in the following). We write:\vspace{-0.1cm}
\begin{equation}\vspace{-0.1cm}
	\hat{\mathbf{v}}_i = \mathbf{v}_i + \boldsymbol{\epsilon}_i,
\end{equation}
where $\boldsymbol{\epsilon}_i$ is a uniformly distributed quantization noise, with $\mathbb{E}[\boldsymbol{\epsilon}_i] = 0$. This is a general and standard model for the quantization noise, widely adopted in the literature, see, e.g.,~\cite{UVeQFed}. 

The aim of the bit allocation is to provide the best possible Hessian approximation given a bit budget. Hence, the quantization scheme design is obtained as the solution of the following problem:\vspace{-0.1cm}
\begin{equation}\label{eq:PBM_General}\vspace{-0.1cm}
	\begin{aligned}
		\min_{b_1, ..., b_q, q} \quad & \ \ \mathbb{E}[\|\mathbf{H} - \hat{\mathbf{H}}_{q, \rho_q}(b_1, ..., b_q)\|_{\mathcal{F}}^2 | \{{\ve}_i,\lambda_i\}_{i = 1}^q]\\
		\textrm{s.t.} \quad & \sum_{i = 1}^q b_i = B\\
		& 0 \leq \ b_i \leq b_{\max}, \forall \, i,    \\
	\end{aligned}
\end{equation} 
where $\hat{\mathbf{H}}_{q, \rho_q}(b_1, ..., b_q)$ is defined in (\ref{eq:HessQDef}). The operator $\|\cdot\|_{\mathcal{F}}$ denotes the Frobenius norm. Note that $q$ is a variable determining the approximation parameter $\rho_q$. The constant $B$ denotes the bit budget, normalized by $n$: denoting the total number of available bits by $B_{\rm tot}$, it holds $B = \floor{B_{\rm tot}/n}$. The integer $b_{\max}$ is the maximum number of bits per vector component. In the following, for ease of notation, we omit the conditioned values from the expectation expression of the squared Frobenius norm introduced in \eq{eq:PBM_General}. For simplicity, we define $\hat{\mathbf{H}}_{q, \rho_q}:=\hat{\mathbf{H}}_{q, \rho_q}(b_1, ..., b_q)$. Denoting by $\tr(\cdot)$ the trace operator, we have that\vspace{-0.1cm}
\begin{equation}\label{eq:defErrHessTr}\vspace{-0.1cm}
	\begin{aligned}
		\mathbb{E}[\|\mathbf{H} - \hat{\mathbf{H}}_{q, \rho_q}\|_{\mathcal{F}}^2] = \mathbb{E}[\tr((\mathbf{H} - \hat{\mathbf{H}}_{q, \rho_q})(\mathbf{H} - \hat{\mathbf{H}}_{q, \rho_q}))],
	\end{aligned}
\end{equation}
where we can write, denoting the unitary eigenvector matrix by $\mathbf{V} := [\ve_1, ..., \ve_n]$,\vspace{-0.2cm}
\begin{equation}\label{eq:HessApproxDiff}\vspace{-0.1cm}
	\begin{aligned}
		\mathbf{H} - \hat{\mathbf{H}}_{q, \rho_q} = \mathbf{V}(\boldsymbol{\Lambda} - \boldsymbol{\Lambda}_{\rho_q})\mathbf{V}^{\top} + \sum_{i = 1}^{q}(\lambda_i - \rho_q)\delta \mathbf{V}_i.
	\end{aligned}
\end{equation}
defining $\delta \mathbf{V}_i:= ({\ve}_i{\ve}_i^{\top} - {\veq}_i{\veq}_i^{\top})$. Plugging (\ref{eq:HessApproxDiff}) into (\ref{eq:defErrHessTr}):\vspace{-0.1cm}
\begin{equation}\label{eq:FrobNorm}\vspace{-0.1cm}
	\begin{aligned}
		&\mathbb{E} [\|\mathbf{H} - \hat{\mathbf{H}}_{q, \rho_q}\|_{\mathcal{F}}^2] = \tr(\mathbf{V}(\boldsymbol{\Lambda} - \boldsymbol{\Lambda}_{\rho_q})^2\mathbf{V}^{\top}) \\
            &+ 2\tr \left (\sum_{i = q+1}^{n}(\bar{\lambda}_{q,i} ){\ve}_i{\ve}_i^{\top}\sum_{i=1}^{q}(\bar{\lambda}_{q,i})\mathbb{E}[\delta\mathbf{V}_i] \right ) \\
            & + \mathbb{E}\left[\tr\left (\sum_{i = 1}^q\bar{\lambda}_{q,i}^2\delta \mathbf{V}_i\delta\mathbf{V}_i \right) + \tr\left(\sum_{\substack{i, j = 1 \\ i\neq j}}^q\bar{\lambda}_{q,i} \bar{\lambda}_{q,j}\delta \mathbf{V}_i\delta \mathbf{V}_j\right)\right].
	\end{aligned}
\end{equation}
where $\bar{\lambda}_{q, i} := \lambda_i - \rho_q$.
The first term of the previous expression does not depend on the quantization strategy, but only on the choice of $q$. The second and third terms, instead, both depend on $q$ and on the quantization strategy through the matrices $\{\delta\mathbf{V}_i\}_{i = 1}^q$. In the next section, we consider the special case of scalar uniform quantization of the eigenvectors' coordinates.\vspace{-0.2cm}
\subsection{Scalar uniform quantization}\label{sec:scalarUniform}\vspace{-0.1cm}
In the case of scalar uniform quantization, each component of vector $\mathbf{v}_i$ is uniformly quantized in the range $[-1, 1]$. Applying dithering, the quantization error vector has i.i.d. uniformly distributed components of known covariance~\cite{UVeQFed}. We can write
\begin{equation}\label{eq:defConversionBits}
	\mathbb{E}[\boldsymbol{\epsilon}_i\boldsymbol{\epsilon}_i^\top] = \sigma_i^2 I, \ \textrm{with}\ \ \sigma_i^2 = \mathbb{E}[\boldsymbol{\epsilon}_{ij}^2] =  \Delta_i^2/12, \ \Delta_i = 2^{-(b_i-1)}
\end{equation}
with $\Delta_i$ being the quantization interval length, and $b_i$ the number of bits assigned to each coordinate of the $i$-th eigenvector. 
After some algebra, we can get\vspace{-0.1cm}
\begin{equation}\vspace{-0.1cm}
	\begin{aligned}
		\mathbb{E}[\tr(\delta \mathbf{V}_i)(\delta \mathbf{V}_i)] = \Delta_i^2(a_1(n) + a_2(n)\Delta_i^2),
	\end{aligned}
\end{equation}
using the fact that $\alpha_i^4 = \Delta_i^4/80 = \mathbb{E}[\epsilon_{ij}^4]$, and defining $a_1(n) := \frac{1}{12} + \frac{n}{6}, \ a_2(n) := \frac{n}{80} + \frac{n(n-1)}{12^2}$. 
With similar calculations, one gets\vspace{-0.1cm} 
\begin{equation}\vspace{-0.1cm}
	\mathbb{E}[\tr(\delta \mathbf{V}_i \delta \mathbf{V}_j)]  = n\sigma_i^2\sigma_j^2 = \frac{n\Delta_i^{2}\Delta_j^2}{12^2} = a_3(n)\Delta_i^2\Delta_j^2,
\end{equation}
with $a_3(n) := \frac{n}{12^2}$.
The expectation of the Frobenius norm of the quantization error in \eq{eq:FrobNorm} can then be written as\vspace{-0.1cm}
\begin{equation}\label{eq:completeError}\vspace{-0.1cm}
	\begin{aligned}
		&\mathbb{E}[\|\mathbf{H} - \hat{\mathbf{H}}_{q, \rho_q}\|_{\mathcal{F}}^2]  = \sum_{i = q+1}^{n}\bar{\lambda}_{q, i}^2 + d_q\sum_{i = 1}^{q}\bar{\lambda}_{q, i}\Delta_i^2 \\&+ \sum_{i = 1}^{q}\bar{\lambda}_{q, i}^2\Delta_i^2(a_1(n) + a_2(n)\Delta_i^2) + \sum_{\substack{i,j = 1\\ i\neq j}}^q\bar{\lambda}_{q, i}\bar{\lambda}_{q,j}a_3(n)\Delta_i^2\Delta_j^2,
	\end{aligned}
\end{equation}
with $d_q = \frac{1}{6}(\sum_{i= q+1}^{n}(\rho_q - \lambda_i))$.
Our objective is to pick the integer parameter $q$ and the quantization intervals $\Delta_1, ..., \Delta_q$ so as to minimize (\ref{eq:defErrHessTr}), with the constraint that $\sum_{i = 1}^q b_i = B$, with $B = \floor{B_{\rm tot}/n}$, where $B_{\rm tot}$ is the number of available bits. Given that $b_i = -\log \Delta_i + 1$, we see that the constraint becomes $\sum_{i = 1}^q \log \Delta_i = q-B$, which is equivalent to $\sum_{i = 1}^q \log \Delta_i^2 = 2(q-B)$. Defining $x_i:=\Delta_i^2$ and $\mathbf{x}_q = (x_1, ..., x_q)$, we define the expectation of the quantization error as a cost function $f$:\vspace{-0.1cm}
\begin{equation}\vspace{-0.1cm}
	f(\mathbf{x}_q, q):= \mathbb{E}[\|\mathbf{H} - \hat{\mathbf{H}}_{q, \rho_q}\|_{\mathcal{F}}^2],
\end{equation}
and we aim to minimize such cost function over the choice of $q$ and over the choice of $\mathbf{x}_q$. We can rewrite \eq{eq:completeError} as\vspace{-0.1cm}
\begin{equation*}\vspace{-0.1cm}
	\begin{aligned}
		f(\mathbf{x}_q, q) & = \sum_{i = q+1}^{n}\bar{\lambda}_{q, i}^2 + \sum_{i = 1}^{q}\gamma_{n,q, i}x_i + a_2(n)\sum_{i = 1}^{q}\bar{\lambda}_{q,i}^2x_i^2 \\&+ a_3(n)\sum_{\substack{i,j = 1\\i\neq j}}^{q} \bar{\lambda}_{q,i}\bar{\lambda}_{q,j}x_ix_j,
	\end{aligned}
\end{equation*}
where $\gamma_{n,q, i} :=  d_q\bar{\lambda}_{q, i} + a_1(n)\bar{\lambda}_{q, i}^2$. The optimization problem is thus turned into the following equivalent form:\vspace{-0.1cm}
\begin{equation}\label{eq:optPbm}\vspace{-0.1cm}
	\begin{aligned}
		\min_{\mathbf{x}_q, q} \quad & f(\mathbf{x}_q, q)\\
		\textrm{s.t.} \quad & -\sum_{i = 1}^q \log x_i \leq 2(B-q)\\
		& 0 < \ x_i \leq 4, \  i = 1, ..., q    \\
	\end{aligned}
\end{equation}
where the last constraint ($x_i \leq 4$) amounts to requiring $b_i \geq 0, i = 1, ..., q$. At optimality, the constraint \mbox{$-\sum_{i = 1}^q \log x_i \leq 2(B-q)$} will be satisfied with equality. The solution to the optimization problem (\ref{eq:optPbm}) needs to be converted in a vector of bits. This can be done by converting each $x_i$ back to $b_i$ using (\ref{eq:defConversionBits}) and then rounding each $b_i$ to the closest integer, being careful to meet the bit budget $\sum_{i = 1}^q b_i = B$.
\newtheorem{lemma}[theorem]{Lemma}
\newtheorem{remark}[theorem]{Remark}
\newtheorem{corollary}[theorem]{Corollary}
\theoremstyle{definition}
\begin{lemma}\label{lemma:Convex}\vspace{-0.15cm}
	For any $q = 1,...,n$, the cost function $f(\mathbf{x}_q, q)$ is strictly convex in $\mathbf{x}_q = (x_1, \dots, x_q)^\top$.
	\begin{proof} Let $\bar{\boldsymbol{\lambda}}_q:=(\bar{\lambda}_1, ..., \bar{\lambda}_q)^\top$, ${\boldsymbol{\gamma}_{n, q}}:=(\gamma_{n, q, 1}, ..., \gamma_{n, q, q})^\top$, $\bar{\boldsymbol{\Lambda}}_q:=\diag(\bar{\lambda}_1^2, ..., \bar{\lambda}_q^2)$,
 and $ \bar{\boldsymbol{\Lambda}}_c \in \mathbb{R}^{q\times q}$ a matrix such that $(\bar{\boldsymbol{\Lambda}}_c)_{i, j} = \bar{\lambda}_{i}\bar{\lambda}_j(1 - \delta_{ij})$, where $\delta_{ii} = 1$ and $\delta_{ij} = 0$ for $i\neq j$. Note that $a_2(n) = \frac{n}{80} + \frac{n(n-1)}{12^2} >  \frac{n}{12^2}= a_3(n)$. Omitting terms that do not depend on $\mathbf{x}_q$, the cost can be rewritten as
	\begin{equation}
		\begin{aligned}
			f(\mathbf{x}_q, q)&= \boldsymbol{\gamma}_{n, q}^{\top} \mathbf{x}_q + a_2(n)\mathbf{x}_q^\top\bar{\boldsymbol{\Lambda}}_q\mathbf{x}_q + a_3(n)\mathbf{x}_q^\top\bar{\boldsymbol{\Lambda}}_c\mathbf{x}_q \\&= \boldsymbol{\gamma}_{n, q}^\top \mathbf{x}_q + \mathbf{x}_q^\top (a_3(n)\bar{\boldsymbol{\Lambda}}_q + (a_2(n) - a_3(n))\bar{\boldsymbol{\lambda}}_q\bar{\boldsymbol{\lambda}}_q^\top) \mathbf{x}_q \\&= \boldsymbol{\gamma}_{n, q}^\top \mathbf{x}_q + \mathbf{x}_q^\top\mathbf{A}_q \mathbf{x}_q,
		\end{aligned}
	\end{equation}
	and because of the fact that $a_2(n) > a_3(n)$, we have
	\begin{equation}
		\mathbf{A}_q = a_3(n)\bar{\boldsymbol{\Lambda}}_q + (a_2(n) - a_3(n))\bar{\boldsymbol{\lambda}}_q\bar{\boldsymbol{\lambda}}_q^\top > 0.
	\end{equation} 
\end{proof}
\end{lemma}\vspace{-0.2cm}
Given the convexity of the constraints, and the strict convexity of the objective function $f(\mathbf{x}_q)$ for any $q = 1, ..., n$ the optimization problem can be solved by solving $n$ convex problems whose solution $\mathbf{x}_q^{*}$ is unique. The optimal solution can be found as the tuple $\{\mathbf{x}_{q^*}^*, q^*\}$, with $q^* = \arg\!\min_{q}\{f(\mathbf{x}_q^{*}, q)\}$.

\section{Q-SHED: algorithm design}
SHED~\cite{NDF_Newton} is designed to make use of Hessian approximations obtained with few Hessian EEPs. In~\cite{NDF_Newton}, it has been shown that incrementally (per iteration) transmitting additional EEPs improves the converges rate. In this section, we augment SHED with the optimal bit allocation of the previous section, making it suitable to incrementally refine the Hessian approximation at the aggregator. The full technique is illustrated in Algorithm~\ref{alg:2}, and the details are provided in the following sections.\vspace{-0.07cm}
\subsection{Uniform scalar quantization with incremental refinements}\vspace{-0.05cm}
Let $\mathbf{H}(\boldsymbol{\theta}^{k_t})$ be the Hessian computed for parameter $\boldsymbol{\theta}^{k_t}$ at round $k_t$. At each round $t \geq k_t$, a number of bits $B_t$ is sent to represent second-order information. At each round, we use newly available bits to incrementally refine the approximation of $\mathbf{H}(\boldsymbol{\theta}^{k_t})$. From now on, eigenvectors are always denoted by $\mathbf{v}_i = \mathbf{v}_i(\boldsymbol{\theta}^{k_t})$, i.e., they are always the eigenvectors of the most recently computed (and possibly outdated) Hessian. If $t = k_t$, the optimal bit allocation for eigenvectors $\mathbf{v}_1, ..., \mathbf{v}_n$ is provided by the scheme presented in Sec.~\ref{sec:scalarUniform}. Fix $t > k_t$. Let $b_{t-1}(i)$ denote the bits allocated to each coordinate of eigenvector $\mathbf{v}_i$ up to round $t-1$, and let $b_{i, t}$ be the number of bits to be used together with $b_{t-1}(i)$, at round $t$, to refine the approximation of the coordinates of $\mathbf{v}_i$. We can write \vspace{-0.2cm}
\begin{equation}\label{eq:quantRef}
	b_t(i) := b_{t-1}(i) + b_{i, t}, \ \ \Delta_{t, i} := \frac{2}{2^{b_t(i)}} := 2^{-b_{t-1}(i)}2^{-b_{i, t}+1}
\end{equation}
with $b_t(i)$ the number of bits sent up to round $t$. The interval $\Delta_{t, i}$ is the quantization interval resulting from adding $b_{i, t}$ bits for the refinement of the $i$-th eigenvector information, for which $b_i^{(t-1)}$ had been previously allocated. We can plug these intervals into \eq{eq:completeError}, and defining $x_{t, i} := 2^{-2(b_{i, t}-1)}$,  $\Tilde{\gamma}_{n, q_{t}, i} := 2^{-2b_{t-1}(i)}{\gamma}_{n, q_{t}, i}$,  $\Tilde{\lambda}_{t, q_{t},i} := 2^{-2b_{t-1}(i)}\bar{\lambda}_{q_{t}, i}$, we get a cost $f(\mathbf{x}_{q_{t}}, q_{t})$, with $\mathbf{x}_{q_{t}} = (x_{t, 1}, ..., x_{t, q_{t}})$,\vspace{-0.15cm}
\begin{equation}\vspace{-0.15cm}
	\begin{aligned}
		f(\mathbf{x}_{q_{t}}, {q_{t}}) &= \sum_{i = {q_{t}}+1}^{n}\bar{\lambda}_{{q_{t}}, i}^2 +\sum_{i = 1}^{{q_{t}}}\Tilde{\gamma}_{n, {q_{t}}, i}x_{t,i} \\+ b(n)\sum_{i = 1}^{{q_{t}}}&\Tilde{\lambda}_{t, {q_{t}}, i}^2x_{t,i}^2 + c(n)\sum_{\substack{i,j = 1\\ i\neq j}}^{q_{t}}\Tilde{\lambda}_{t, {q_{t}}, i}\Tilde{\lambda}_{t, {q_{t}},j}x_{t,i}x_{t,j}.
	\end{aligned}
\end{equation}
Following the same proof technique as for Lemma \ref{lemma:Convex}, it can be shown that the cost $f(\mathbf{x}_{q_{t}}, q_{t})$ is strictly convex in $\mathbf{x}_{q_{t}}$ for any ${q_{t}} = 1, ..., n$. Given that up to round $t-1$, $q_{t -1}$ eigenvectors were considered for bit allocation, it is easy to see that it needs to be $q_{t} \geq q_{t -1}$. Similarly to Sec.~\ref{sec:scalarUniform}, we formulate the optimal bit allocation of bits $\{b_{i, t}\}_{i = 1}^{q_{t}}$ as 
\begin{equation}\label{eq:optPbmIncr}\vspace{-0.1cm}
	\begin{aligned}
		\min_{\mathbf{x}_{{q_{t}}}, {q_{t}} \geq q_{t-1}} \quad & f(\mathbf{x}_{q_{t}}, {q_{t}})\\
		\textrm{s.t.} \quad & -\sum_{i = 1}^{q_{t}} \log x_{t, i} \leq 2(B_{t}-{q_{t}})\\
		& 0 < \ x_{t, i} \leq 4, \  i = 1, ..., {q_{t}}.    \\
	\end{aligned}
\end{equation}
The problem can be solved by finding the unique solution to the $n-q_{t-1} + 1$ strictly convex problems corresponding to the different choices of $q_{t} = q_{t-1}, q_{t-1}+1, ..., n$. As before, the solution to problem (\ref{eq:optPbmIncr}) needs to be converted to integer numbers, for example by rounding the corresponding allocated number of bits to the closest integer, being careful to retain $\sum_{i = 1}^q b_{i, t} = B_t$. Sorting the eigenvalues in a decreasing order, we get a monotonically decreasing sequence of allocated bits to the corresponding eigenvectors. To provide an example, with the FMNIST dataset (see Sec. \ref{sec:results}), at a certain iteration $t$ of the incremental algorithm, an agent allocates bits $b_t = [3, 3, 2, 2, 2, 1, 1, 1, 1, 1, 1]$ to the first $11$ eigenvectors, whose corresponding (rounded) eigenvalues are $[0.21, 0.11, 0.06, 0.03, 0.03, 0.02, 0.02, 0.01, 0.01, 0.01, 0.01]$.\vspace{-0.1cm}
\subsection{Multi-agent setting: notation and definitions}\vspace{-0.1cm}
To illustrate the integration of our incremental quantization scheme with SHED, we introduce some definitions for the multi-agent setting. We denote by $B_{t}^{(d)}$ the bit-budget of device $d$ at iteration $t$. Let  $\rho_{t}^{(d)} = \lambda_{q_t^{(d)}+1, t}^{(d)}$ be the Hessian approximation parameter of device $d$ at iteration $t$, function of the $q_t^{(d)}$-th eigenvalue of the $d$-th device, where the integer $q_t^{(d)}$ is tuned by device $d$ as part of the bit-allocation scheme at iteration $t$. Let $\mathbf{g}_t^{(d)}$, $\mathbf{H}_t^{(d)}$, $\hat{\mathbf{H}}_t^{(d)}$ be the gradient, Hessian, and Hessian approximation, respectively, of device $d$. We denote by $\mathbf{v}_i^{(d)}$ and $\hat{\mathbf{v}}_i^{(d)}$ the $i$-th eigenvector of the $d$-th device and its quantized version, respectively. Note that eigenvectors always correspond to the last computed Hessian $\mathbf{H}(\boldsymbol{\theta}^{k_t})$, with $k_t \leq t$. The integer $b_t^{(d)}(q)$ denotes the number of bits allocated by device $d$ to the $q$-th eigenvector coordinates up to iteration $t$, while $b_{q,t}^{(d)}$ is the per-iteration bits allocated to the $q$-th eigenvector, i.e., $b_t^{(d)}(q) = b_{t-1}^{(d)}(q) + b_{q, t}^{(d)}$. We define $\mathcal{A}$ to be the set of devices involved in the optimization, $\mathcal{I}$ the set of iteration indices in which each device recomputes its local Hessian, $f^{(d)}$ the cost function of device $d$, and $\epsilon > 0$ the gradient norm threshold. Hessian approximations are built at the aggregator in the following way:\vspace{-0.2cm}
\begin{equation}\label{eq:HessApprox}\vspace{-0.12cm}
	\hat{\mathbf{H}}_t = \frac{1}{N}\sum_{d = 1}^{M}N_d\hat{\mathbf{H}}_t^{(d)}, \ \ \hat{\mathbf{H}}_t^{(d)} = \sum_{i = 1}^{q_t^{(d)}}\bar{\lambda}_i^{(d)}\hat{\mathbf{v}}_i^{(d)}\hat{\mathbf{v}}_i^{(d)\top} + \rho_t^{(d)}\mathbf{I},
\end{equation}
where $\bar{\lambda}_i^{(d)} = \lambda_i^{(d)} - \rho_t^{(d)}$. Incremental quantization allows devices to refine the previously transmitted quantized version of their eigenvectors by adding information bits, see (\ref{eq:quantRef}). We denote the set of information bits of device $d$ sent to quantize or refine previously sent quantized eigenvectors by $Q_t^{(d)}$.\vspace{-0.1cm}
\subsection{Heuristic choice of $q_{t}^{(d)}$}\vspace{-0.1cm}
To reduce the computational burden at the edge devices and to solve the bit-allocation problem only once per round, we propose a heuristic strategy for each device to choose $q^{(d)}_t$: at each incremental round $t$, instead of inspecting all the options corresponding to $q_{t-1}^{(d)}, ..., q_n^{(d)}$, which would provide the exact solution, but would require solving problem (\ref{eq:optPbmIncr}) $n-q_{t-1}^{(d)} + 1$ times. We fix $\bar{q} = q_{t-1}^{(d)} + B_{t}^{(d)}$: With this choice of $\bar{q}$, we solve problem (\ref{eq:optPbmIncr}), and we subsequently convert the solution to bits obtaining $\{b_{i, t}^{(d)}\}_{i = 1}^{\bar{q}}$ and $\{b_{t}^{(d)}(i)\}_{i = 1}^{\bar{q}}$. We then fix the value
\begin{equation}\label{eq:heurChoiceQ}
	q_{t}^{(d)} = \hat{q}_t^{(d)}(\{b_{t}^{(d)}(i)\}_{i = 1}^{\bar{q}}) := \max_q\{q: b_{t}^{(d)}(q) > 0\}
\end{equation}
\begin{algorithm}[b!]
	\begin{algorithmic}[1]
		\Require{$\{f^{(d)}\}_{d = 1}^{M}$, $\mathcal{I}$, $\boldsymbol{\theta}^1$, $f$, $\nabla f({\x}^1)$, $\mathcal{A}$, $\epsilon > 0$}
		\Ensure{$\x^{t}$}
		\State $t\gets 1$
		\While {$\|\nabla f ({\x}^t)\|_2 \geq \epsilon$}
		\For{$device \ d \in \mathcal{A}$}
		\When{received $\x^t$ from the aggregator}
		\If{$t \in \mathcal{I}$} 
		\State $k_t \gets t$ 
		\State compute $\mathbf{H}_t^{(d)} = \nabla^2 f^{(d)}(\boldsymbol{\theta}^t)$ \Comment{renewal}
		\State$\{({\lambda}_{j, t}^{(d)}, {\mathbf{v}}_{j}^{(d)})\}_{j = 1}^{n} \gets \text{eigendecomp}(\mathbf{H}_t^{(i)})$
		\State $q_{t-1}^{(d)} \gets 0$\EndIf
		\State compute $\mathbf{g}_t^{(d)} = \mathbf{g}^{(d)}(\boldsymbol{\theta}^t)= \nabla f^{(d)}(\boldsymbol{\theta}^t)$
		\State $\bar{q} \gets q_{t-1}^{(d)} + B_t^{(d)}$
		\State $\mathbf{x}_{\bar{q}}^* \gets$ \text{solve (\ref{eq:optPbmIncr}) for $q_t = \bar{q}$} with budget $B_t^{(d)}$  
		\State $\{b_{i, t}^{(d)}\}_{i = 1}^{\bar{q}} \gets \text{convertToBits}(\mathbf{x}_{\bar{q}}^*)$ \Comment{back to bits}
		\State $q_t^{(d)} \gets \hat{q}_t^{(d)}(\{b_{t}^{(d)}(i)\}_{i = 1}^{\bar{q}})$ \Comment{see (\ref{eq:heurChoiceQ})}
		\State $\rho_t^{(d)} \gets {\lambda}_{q_t^{(d)}+1, t}^{(d)}$ 
		\State $Q_t^{(d)} \gets \text{quantize}(\{\mathbf{v}_i^{(d)}\}_{i = 1}^{q_{t}^{(d)}}, \{b_{i, t}^{(d)}\}_{i = 1}^{{q}_t^{(d)}}, \{b_{t}^{(d)}(i)\}_{i = 1}^{{q}_t^{(d)}})$\Comment{quantize or refine quantization, see (\ref{eq:quantRef})}
		\State$U_t^{(d)} \gets \{Q_t^{(d)}, \{{\lambda}_{j, t}^{(d)}\}_{j = q^{(d)}_{t-1}+1}^{q^{(d)}_t}, \mathbf{g}_t^{(d)}, {{\rho}}_t^{(d)}\}$
		\State send $U_t^{(d)}$ to the aggregator\EndWhen\EndFor
		\State\hrulefill
		\State at the aggregator:
		\When{received $U_t^{(d)}$ from all devices}
		\State compute $\hat{\mathbf{H}}_t^{(d)},\ \forall{d}$\Comment{see (\ref{eq:HessApprox})}
		\State compute $\hat{\mathbf{H}}_t$ (see (\ref{eq:HessApprox})) and
		$\mathbf{g}_t$
		\State get $\eta_t$ via distributed backtracking line search.
		\State perform Newton-type update (\ref{eq:qN-update})
		\State broadcast $\boldsymbol{\theta}^{t+1}$ to all devices.\EndWhen
		\State$t \gets t + 1$
		\EndWhile
	\end{algorithmic}
	\caption{Q-SHED}\label{alg:2}
\end{algorithm}
\subsection{Convergence analysis}\vspace{-0.1cm}
The choice of Hessian approximation is positive definite by design (see (\ref{eq:HessApprox})). Hence, the algorithm always provides a descent direction and, with a backtracking strategy like in~\cite{GIANT} and \cite{NDF_Newton}, convergence is guaranteed (see Theorem 4 of \cite{NDF_Newton}). Empirical results suggest that linear and superlinear convergence of the original SHED may still be guaranteed under some careful quantization design choices. We leave the analysis of the convergence rate as future work, but we provide an intuition on the convergence rate in the least squares case. In the least squares case, for a given choice of $q$ and of the allocated bits $\{b_{i, t}^{(d)}\}_{i = 1}^{{q}}$ of each device $d$, an easy extension of Theorem~3 in \cite{NDF_Newton} provides the following bound\vspace{-0.1cm}
\begin{equation}
	\|{\x}^{t+1} - {\x}^*\| \leq \kappa_{t}\|\boldsymbol{\theta}^t - \boldsymbol{\theta}^*\|,
\end{equation}
with $\kappa_t = (1 - {(\bar{\lambda}_n - e_t)}/{\bar{\rho}_t})$, where 
$$\bar{\lambda}_n = \frac{1}{N}\sum_{d = 1}^{M}N_d\lambda_n^{(d)} \ \ \bar{\rho}_t = \frac{1}{N}\sum_{d = 1}^{M}N_d\rho_t^{(d)}$$ and
\begin{equation}
	e_t = \frac{1}{N}\sum_{d = 1}^MN_d\sum_{i = 1}^{q_t^{(d)}}(\lambda_i^{(d)}-{\rho}_t^{(d)})\|\delta \mathbf{V}_i^{(d)}\| 
\end{equation}
where $\delta \mathbf{V}_i^{(d)}:= ({\ve}_i^{(d)}{\ve}_i^{(d)\top} - {\veq}_i^{(d)}{\veq}_i^{(d)\top})$. It can be noted how for a sufficiently small quantization error, which can always be achieved by incremental refinements, the convergence rate in the least squares case is at least linear. The extension to the general case is left as a future work.
\section{Empirical Results}\label{sec:results}
\begin{figure}
\centering	\includegraphics[width=0.9\columnwidth, trim = {0, 0, 0, 0}, clip]{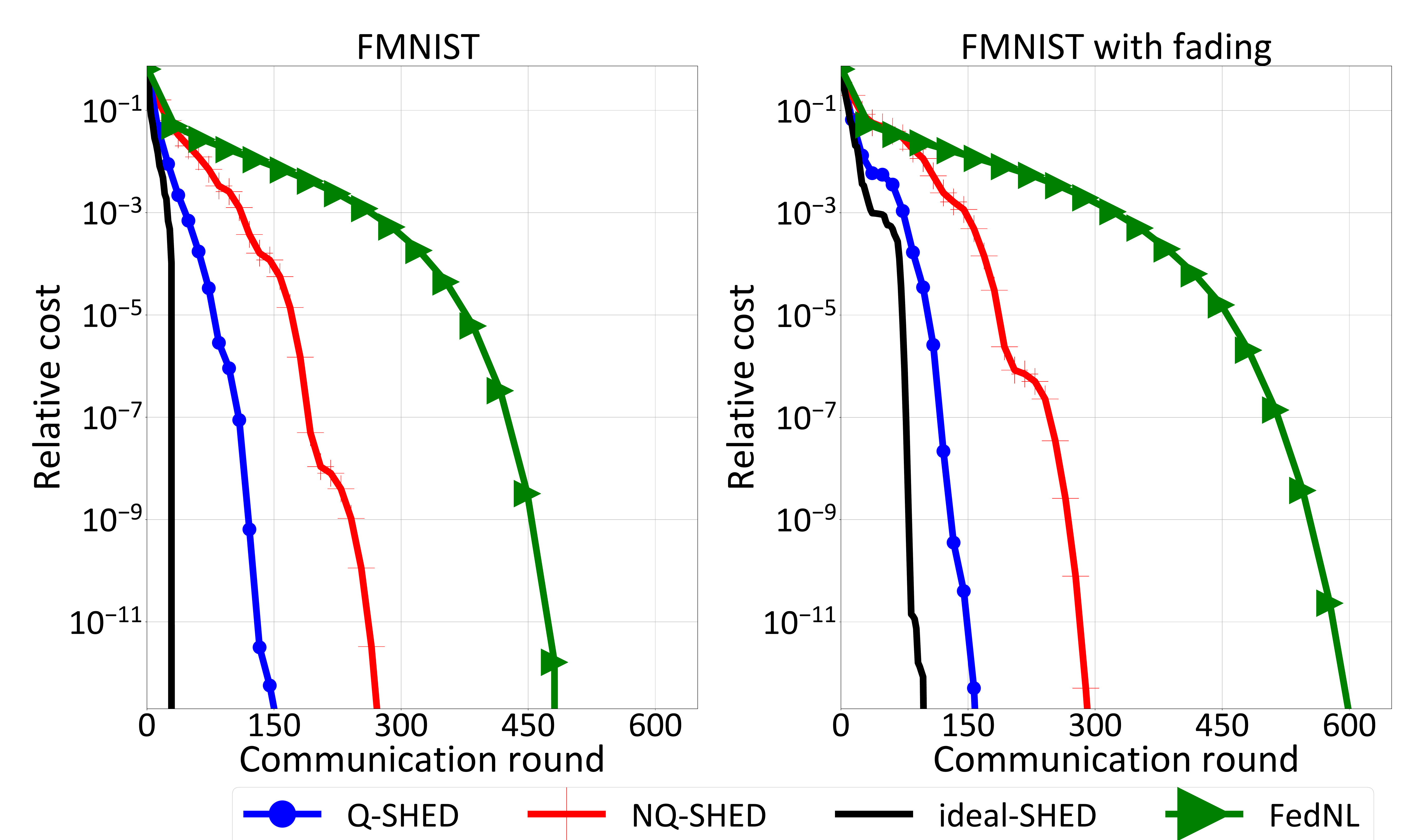}
	\caption{Comparison of Q-SHED against NQ-SHED and FedNL with the FMNIST dataset. With the exception of ideal-SHED, for a fair comparison, in each communication round the algorithms use the same number of bits. Relative cost is $f(\boldsymbol{\theta}^t) - f(\boldsymbol{\theta}^*)$.}
	\label{fig:1}
\end{figure}

\begin{figure}
\centering	\includegraphics[width=0.9\columnwidth, trim = {0, 0, 0, 0}, clip]{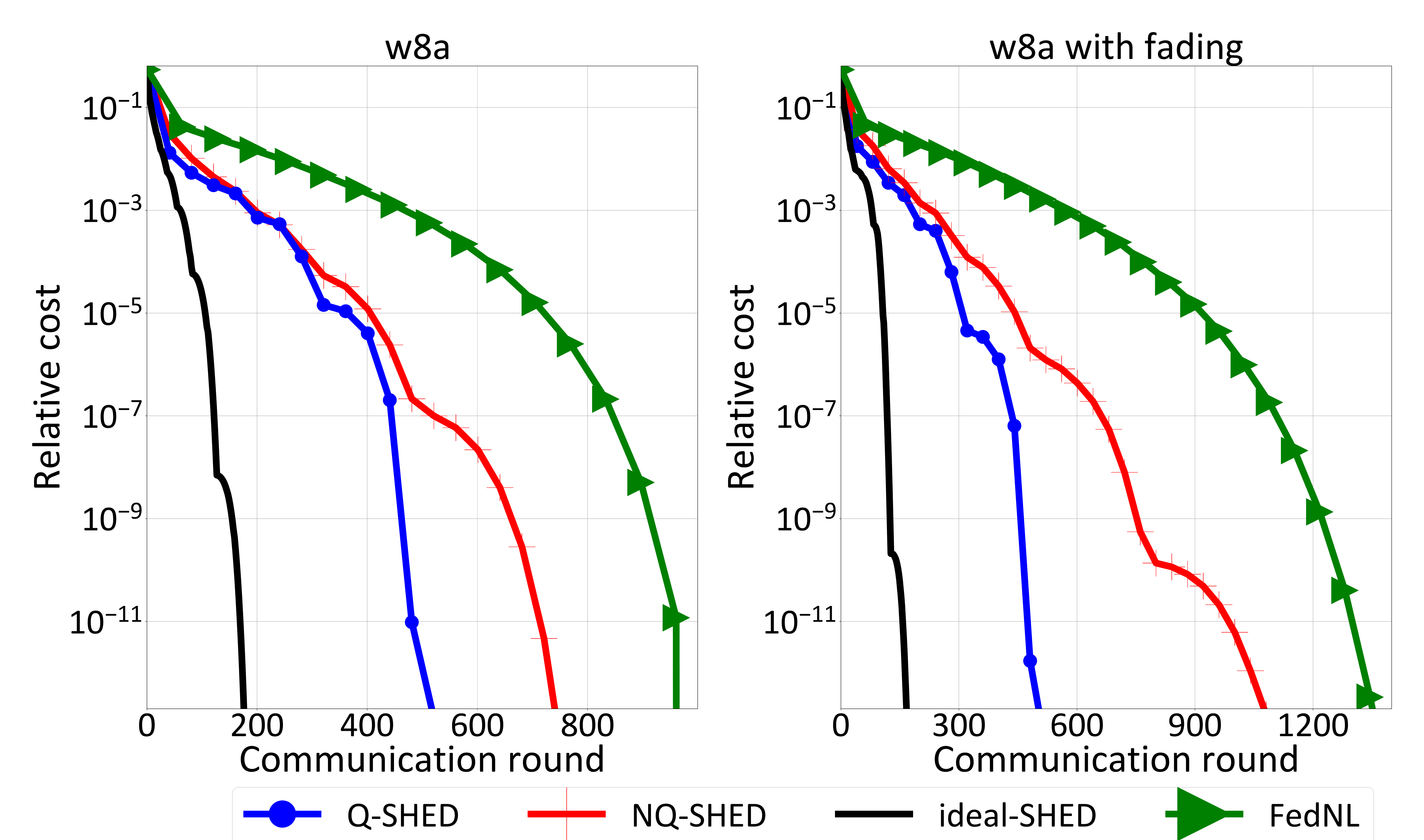}
	\caption{Comparison of Q-SHED against NQ-SHED and FedNL with the w8a dataset. With the exception of ideal-SHED, for a fair comparison, in each communication round the algorithms use the same number of bits. Relative cost is $f(\boldsymbol{\theta}^t) - f(\boldsymbol{\theta}^*)$.}
\label{fig:2}
\end{figure}

In this section, we provide empirical results obtained with two datasets, FMNIST \cite{FMNIST} and w8a \cite{LIBSVM}. We simulate two configurations for the network: one where every device has the same transmission rate at each communication round, and one where the rate changes randomly for each device based on the widely adopted Rayleigh fading model~\cite{Pase, ChannelUncertainty}. For both FMNIST and w8a we build up a binary classification setting with logistic regression (in FMNIST we learn to distinguish class ‘1’ from all the others), simulating a scenario with $M = 8$ devices, each with $500$ data samples. We use L2 regularization with parameter $\mu = 10^{-5}$. For FMNIST, we apply PCA~\cite{PCA-MNIST} to the data to reduce the dimensionality to $n = 90$, while for w8a we keep the original data dimensionality, $n = 300$. To simulate the fading channels, we adopt the following simple model. We consider that all the devices allocate the same bandwidth $\beta$ for the communication with the aggregator and write the achievable transmission rate as (see, e.g.,~\cite{Pase, ChannelUncertainty})
\begin{equation}
	R^{(d)} = \beta\log_2(1 + \gamma\Gamma^{(d)})
\end{equation}
where $\Gamma^{(d)}$ is a value related to transmission power and environmental attenuation for user $d$. For simplicity, we fix $\Gamma^{(d)} = \Gamma = 1$ for all users (in \cite{Pase}, for instance, $\Gamma = 1$ and $\Gamma = 10$ were considered). The only source of variability is then $\gamma \sim \exp(\nu)$, modelling the Rayleigh fading effect. We fix $\nu = 1$. Specifically, to simulate the different bit budgets, we compute the individual bit budget of each device as $B_t^{(d)} = B\log_2(1 + \gamma\Gamma^{(d)})$, setting $B = 2b_{\max}$. We fix $b_{\max} = 16$. In the non-fading case, the bit budget for each device is constant and set to $B_t^{(d)} = 2b_{\max}$. We consider a scenario where the full-quality gradient is always transmitted to the aggregator by the devices. We compare Q-SHED against an ideal version of SHED, dubbed ideal-SHED, where the eigenvectors that are quantized by Q-SHED are transmitted at full quality. We also compare Q-SHED against a naively-quantized counterpart, NQ-SHED, for which all bits are allocated to the first eigenvectors, and the state-of-the-art FedNL~\cite{FedNL} with rank-1 compressors. With the exception of ideal-SHED, the per-round bit budget of the considered algorithms is the same. We have experimented with the possibility of quantizing the second-order information of FedNL, but we observed a performance degradation. Hence, when the bit budget of a device is not enough for communicating the \mbox{rank-1} compression of the Hessian drift at full quality, we only use the device's local gradient. We do the same for NQ-SHED. The results on FMNIST and w8a are shown in Figs.~\ref{fig:1} and \ref{fig:2}, respectively. In both cases, it is possible to appreciate the robustness of Q-SHED in terms of iterations required for convergence, while both NQ-SHED and FedNL performance is degraded in the presence of fading channels. In terms of convergence speed, the results show that Q-SHED provides performance improvements against the selected competing solutions between 30\% and 60\%. 

\vspace{0.1cm}
\section{Conclusion and future work}
We have empirically shown that Q-SHED outperforms its naively-quantized version as well as state-of-the-art algorithms like FedNL. Future works include an in-depth analysis of the convergence rate, and the adoption of more advanced quantization schemes, like vector quantization techniques.
\section{Acknowledgment}
This work has been supported, in part, by the Italian Ministry of Education, University and Research, through the PRIN project no. 2017NS9FEY and by the European Union under the Italian National Recovery and Resilience Plan (NRRP) of NextGenerationEU, partnership on “Telecommunications of the Future” (PE0000001 - program
“RESTART”).
.





\bibliographystyle{IEEEtran}	\bibliography{IEEEabrv,biblio}
\end{document}